\documentclass[twoside,final,1p,times]{elsarticle}
\usepackage{amsmath}
\usepackage{subfigure}
\usepackage{multirow}
\usepackage{url}
\usepackage{fancyhdr}
\usepackage{tikz}
\usepackage{color}
\journal{Data Intelligence}
\usepackage{booktabs}
\newcommand\shorttitle{\textcolor[RGB]{20,180,190}{Data Intelligence}}
\begin{document}
\begin{frontmatter}
{\title{\textbf{Open-Source Full-Duplex Conversational Datasets for Natural and Interactive Speech Synthesis}}}

\author{Zhitong Zhou}
\author{Qingqing Zhang}
\author{Lei Luo}
\author{Jiechen Liu}
\author{Ruohua Zhou\daggerfootnote{Corresponding author: Ruohua Zhou (Email: zhouruohua@bucea.edu.cn; ORCID:)}}

\address{Beijing University of Civil Engineering and Architecture}
\address{No.15 Yongyuan Road, Huangcun Town, Daxing District, Beijing 102616, China}
\address{Magic Data}
\address{5th floor, Building G, No. 44 BeiSanHuan Middle Road, Haidian District, Beijing 100088, China}

\begin{abstract}
Full-duplex, spontaneous conversational data are essential for enhancing the naturalness and interactivity of synthesized speech in conversational TTS systems. 
We present two open-source dual-track conversational speech datasets, one in Chinese and one in English, designed to enhance the naturalness of synthesized speech by providing more realistic conversational data.
The two datasets contain a total of 15 hours of natural, spontaneous conversations recorded in isolated rooms, which produces separate high-quality audio tracks for each speaker.
The conversations cover diverse daily topics and domains, capturing realistic interaction patterns including frequent overlaps, backchannel responses, laughter, and other non-verbal vocalizations. 
We introduce the data collection procedure, transcription and annotation methods. 
We demonstrate the utility of these corpora by fine-tuning a baseline TTS model with the proposed datasets. 
The fine-tuned TTS model achieves higher subjective and objective evaluation metrics compared to the baseline, indicating improved naturalness and conversational realism in synthetic speech. 
All data\footnote{https://magichub.com/datasets/multi-stream-spontaneous-conversation-training-datasets\_chinese/}
\footnote{https://magichub.com/datasets/multi-stream-spontaneous-conversation-training-datasets\_english/}, annotations, and supporting code for fine-tuning and evaluation are made available to facilitate further research in conversational speech synthesis.
\end{abstract}


\end{frontmatter}

\section{Background}
Full-duplex conversation features natural phenomena such as interruptions, overlaps, and spontaneous interjections \cite{1}, which are crucial for improving the naturalness of synthesized speech.
Unlike traditional half-duplex (turn-by-turn) interactions, full-duplex voice interaction enables more natural conversations with fluid backchannels, interruptions, and quicker turn-taking, which features common in human conversation \cite{2}. 
For example, researchers have found that more than 40\% of speaker turns in natural conversation involve some overlapping speech \cite{1}. 
Such overlaps include user interruptions (barge-ins), listener backchannels (e.g. “mm-hmm”, “right” interjections), and other spontaneous reactions \cite{3}. 
Moreover, human speech exhibits dynamic pacing, with turn transitions often occurring within just a few hundred milliseconds, requiring conversational system to flexibly adjust response timing so that interaction feels fluid rather than mechanical \cite{3_1}. 
This underscores the difficulties of training conversational TTS models on data that reflect real dialog dynamics, including simultaneous speech. 

Most existing speech corpora for TTS and conversational modeling are turn-based. 
For example, classic conversational datasets like Switchboard (telephone conversations) captured many conversations but were constrained to sequential turn exchanges with limited overlap \cite{4}. 
Similarly, TTS-specific conversational datasets like DailyTalk containing over 2,500 conversations turns recorded at high fidelity, aim to support contextual speech synthesis but still seldom include spontaneous overlapping speech or annotated interruption dynamics \cite{5}. 
In many such corpora, overlaps and turn transitions are either infrequent or not explicitly annotated, limiting their usefulness for training full-duplex interaction models.

To address this gap, we develop two separate conversational datasets in Chinese and English, each capturing the dynamics of real-time bidirectional speech.
The dataset consists of topic-guided spontaneous conversations between native speakers of Chinese and English, with each speaker recorded independently on a separate mono track. 
This dual-track setup preserves natural overlap and interruption events, resulting in two datasets containing approximately 5 hours of English and 10 hours of Chinese conversational speech, respectively.
Moreover, the dataset captures spontaneous and dynamic conversational behaviors, including backchannels, mid-turn interruptions, and brief overlaps, reflecting interaction patterns typical of real-world conversation. 
We anticipate that these datasets will improve the naturalness of synthesized speech, making it closer to real human conversations.

\section{Methods}

We constructed the full-duplex conversational dataset with high-quality conversational speech in both Chinese and English. 
Two participants, located in separate rooms, use each mobile phone for recording.
The speech was captured as speaker-isolated dual-mono tracks using on-device local recording or a split-track microphone setup.

\subsection{Speaker Recruitment and Prompt Selection}
We recruited native speakers of Mandarin Chinese and English through Magic Data crowdsourcing platform. 
Participants were recruited in pairs, with a preference for dyads who were already familiar with each other (e.g., friends, classmates, or family members). 
Familiarity was prioritized to reduce social inhibition, facilitate spontaneous exchanges, and elicit naturalistic conversational behaviors such as backchannels and interruptions. 
Each pair selected their own discussion themes without restriction, allowing the content to emerge organically from speakers’ shared interests and experiences. 
This open-ended design encouraged expressiveness and variation in discourse structure, prosody, and affect, while avoiding priming effects associated with prescriptive prompts. 
Conversations proceeded freely and were self-paced; no scripted conversation or role assignment was imposed. 
The collection protocol emphasized ecological validity and conversational authenticity, aiming to approximate ordinary human-to-human interaction.
Then, a human-assisted quality control pipeline filtered out recordings with severe noise or hardware artifacts. 

\begin{figure}
    \centering
    \includegraphics[width=0.7\linewidth]{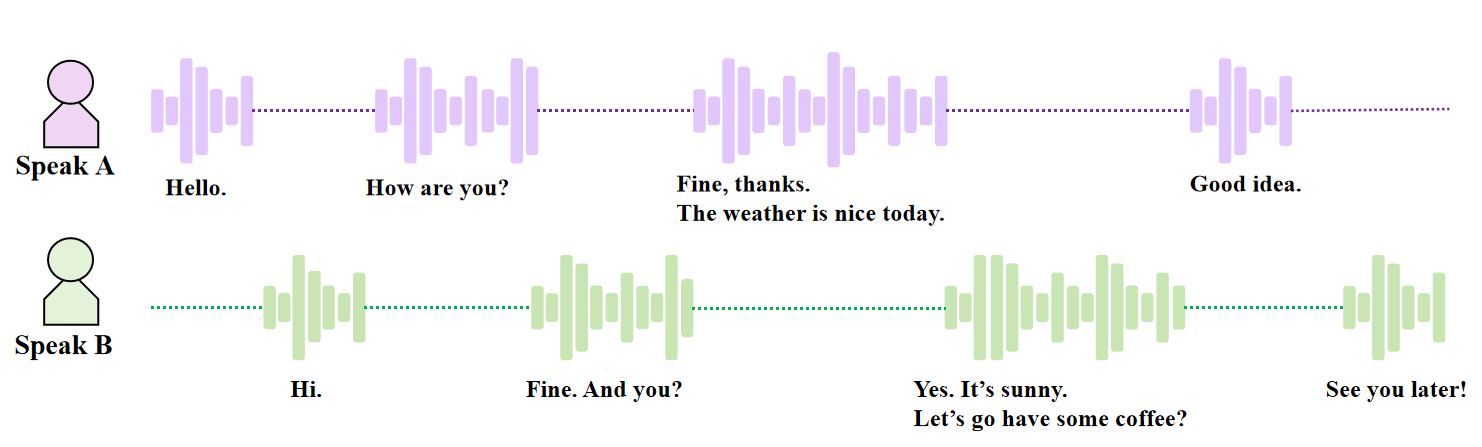}
    \caption{Illustration of Multi-Speaker, Multi-Stream Conversation Recording Format in the Dataset}
    \label{fig:placeholder}
\end{figure}

\subsection{Transcription and Annotation}

Qualified recordings were manually transcribed by trained annotators. 
Each speech was annotated with speaker identity, gender, and precise start and end timestamps.

Concurrent speech was represented as overlapping time intervals on the two speaker tracks. 
In addition to verbatim transcription, the annotation scheme included both paralinguistic and interactional elements. 

Annotators focus not only on character accuracy but also on the correctness of punctuation, sentence segmentation, and contextual elements.
Particularly, preserving semantic integrity is a crucial and essential aspect of the current model training process. We require annotators to perform VAD (Voice Activity Detection) segmentation in accordance with semantic completeness, ensuring that the segmentation reflects semantic integrity. Such annotations are highly scarce and valuable, as they can be utilized not only for training TTS (Text-to-Speech) systems but also for semantic understanding tasks.

\section{Data Records}

The full-duplex conversational dataset as formatted in Figure \ref{fig:placeholder} consists of two language subsets: Chinese and English. 
All recordings are released as dual-track WAV files (16-bit PCM, 16 kHz), with track 1 corresponding to Speaker A and track 2 to Speaker B. 
Each conversation is accompanied by a plain text transcript, where utterances from both speakers 
are merged and ordered chronologically. 
Each line includes the start and end timestamps, speaker ID, gender, and transcribed utterance. 
Both the audio and transcription files are organized per speaker. 
Each transcription file contains timestamped segments for the corresponding speaker’s utterances, with “None” labels marking silent or listening intervals.

File names follow the convention \texttt{A<SessionID>\_S<TopicID>\_0\_G<SpeakerID>}, where `A' indicates group ID, `S' is the topic number, `0' denotes the recording device, and `G' refers to the speaker ID.
The dataset includes a total of 35 recorded conversations (27 in Chinese and 8 in English), spoken by 14 unique speakers. A summary is shown in Table~\ref{tab:dataset_stats}.

\begin{table}[ht]
\centering
\caption{Dataset Statistics}
\label{tab:dataset_stats}
\begin{tabular}{lcccc}
\toprule
\textbf{Language} & \textbf{Conversations} & \textbf{Unique Speakers} & \textbf{Utterance Segments} & \textbf{Total Duration} \\
\midrule
Chinese & 27 & 6 & 2306 & 10 hours \\
English & 8 & 8 & 4149 & 5 hours \\
\bottomrule
\end{tabular}
\end{table}

\section{Technical Validation}

To assess the quality and utility of our full-duplex Chinese–English conversational dataset, we fine-tuned a pretrained CosyVoice‑300M model \cite{9} separately on the Chinese and English datasets and conducted comprehensive objective and subjective evaluations. 
CosyVoice‑300M employs two distinct loss functions corresponding to its modular architecture. 
The text-to-token module is supervised by minimizing the following cross-entropy loss:
\begin{equation}
    \mathcal{L}_{\text{LLM}} = -\frac{1}{L + 1} \sum\nolimits_{l=1}^{L+1} \log q(\mu_l),
\end{equation}
where \( \mu_l \) denotes the $l$-th target token (including EOS), and \( q(\mu_l) \) is the predicted probability.

For the token-to-mel module, an optimal-transport conditional flow matching loss is utilized to learn the distribution of Mel spectrogram and generate samples from it with generated speech tokens as conditions, which is expressed as:
\begin{equation}
    \mathcal{L}_{\text{OT-CFM}} = \mathbb{E}_{t, p_0(X_0), q(X_1)} \left[ \left| \omega_t \left( \phi_t^{\text{OT}}(X_0, X_1) \mid X_1 \right) - \nu_t \left( \phi_t^{\text{OT}}(X_0, X_1) \mid \theta \right) \right| \right],
\end{equation}
where \( \phi_t^{\text{OT}}(X_0, X_1) \) denotes the interpolation along the optimal transport path, \( \omega_t \) is the target vector field, and \( \nu_t \) is the model-predicted vector field conditioned on speaker embedding, semantic tokens, and masked mel-spectrogram features. 
The two modules are trained independently and asynchronously. 
The learning rate and training epochs are set as $10^{-5}$ and 200. 
Chinese utterances were limited to approximately 30 characters, and English utterances to 15 words. 
If a segment fell below the minimum length threshold, it was merged with the subsequent utterance. 

\begin{table}[ht]
\centering
\caption{Objective Metric Improvements After Fine-Tuning (Chinese and English Subsets)}
\label{tab:objective_combined}
\begin{tabular}{llccccc}
\toprule
\textbf{Language} & \textbf{Metric} & \textbf{Original} & \textbf{Fine-tuned} & \textbf{Difference} & \textbf{Improvement} \\
\midrule
\multirow{4}{*}{Chinese}
& Spectrum\_l2 & 28.5452 & 28.0907 & 0.4545 & 1.59\% \\
& F0\_Wasserstein & 63.6259 & 59.1216 & 4.5043 & 7.08\% \\
& ZCR\_Wasserstein & 0.0136 & 0.0134 & 0.0003 & 1.97\% \\
& Energy\_Wasserstein & 0.0104 & 0.0103 & 0.0001 & 0.53\% \\
\midrule
\multirow{4}{*}{English}
& Spectrum\_l2 & 34.5732 & 34.8681 & -0.2949 & -0.85\% \\
& F0\_Wasserstein & 55.8803 & 53.8320 & 2.0482 & 3.67\% \\
& ZCR\_Wasserstein & 0.0074 & 0.0068 & 0.0005 & 7.44\% \\
& Energy\_Wasserstein & 0.0172 & 0.0170 & 0.0002 & 1.33\% \\
\bottomrule
\end{tabular}
\end{table}

\subsection{Objective Evaluation}

We employed four core acoustic metrics to evaluate the similarity between synthesized and real speech:
1) Spectrum\_l2 distance calculates the Euclidean distance between average mel-spectrograms of synthesized and real speech, providing a global measure of spectral (timbre) similarity. 
2) F0 distance quantifies the distributional difference of fundamental frequency (F0), reflecting prosodic features such as intonation and pitch variation; lower values indicate a closer match to natural speech rhythm; 
3) Energy distance measures the discrepancy in loudness dynamics via root-mean-square (RMS) energy distributions, capturing how well the synthesized speech replicates natural amplitude variation; 
4) ZCR distance evaluates the similarity in zero-crossing rate distributions, which relate to voicing patterns, speech rate, and the presence of unvoiced segments.
F0, Energy and ZCR are measured in Wasserstein distance. 
Lower values across these metrics indicate more natural and human-like synthesis performance.

The evaluation revealed consistent improvements across both Chinese and English subsets. 
As summarized in Table~\ref{tab:objective_combined}, the Chinese model exhibited a significant reduction in F0 distance (7.08\%) and mel spectrum distance (1.59\%). 
Similar trends were observed for English, including a 3.67\% reduction in F0 distance and a 7.44\% reduction in ZCR distance, indicating improved prosody and temporal dynamics after fine-tuning.

\begin{figure}[ht]
    \centering
    \includegraphics[width=0.7\linewidth]{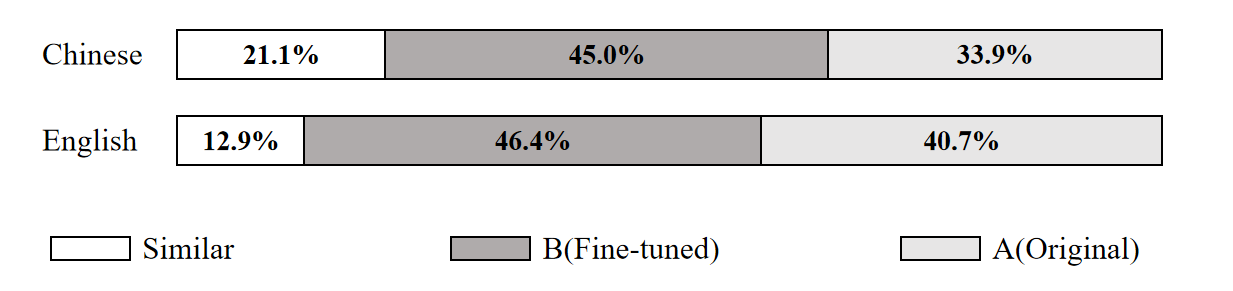}
    \caption{A/B preference distribution}
    \label{fig2}
\end{figure}

\begin{table}[ht]
\centering
\label{T3}
\caption{Comparison of MOS Scores for Chinese and English}
\begin{tabular}{lcc|cc}
\toprule
 & \multicolumn{2}{c}{Chinese} & \multicolumn{2}{c}{English} \\
\cmidrule(lr){2-3} \cmidrule(lr){4-5}
 & Original (A) & Fine-tuned (B) & Original (A) & Fine-tuned (B) \\
\midrule
Naturalness & 4.37 & 4.40 & 4.47 & 4.51 \\
Intelligibility & 4.46 & 4.49 & 4.56 & 4.58 \\
\bottomrule
\end{tabular}
\end{table}

\subsection{Subjective Evaluation}

To further assess perceptual quality, we conducted the following subjective evaluation. 
Native Mandarin Chinese and English listeners (10 participants per language) were recruited to assess the naturalness and intelligibility of the synthesized speech samples. 
In each trial, participants heard two clips generated from the same text prompt, one from the original model (A) and one from the fine-tuned model (B) and judged which sounded more natural.
Each listener evaluated the same 80 pairs of conversation. 
To minimize bias, we adopted a randomized single-blind protocol: the A/B presentation order was shuffled and the model identities were concealed.
We used an A/B preference paradigm (A = original, B = fine-tuned).
The aggregated results are summarized by stacked bars: for Chinese, 21.1\% chose `similar', 45.0\% preferred B and 33.9\% preferred A; for English, 12.9\% chose `similar', 46.4\% preferred B and 40.7\% preferred A. 
Furthermore, as shown in Table 3, complementary MOS ratings further show small but consistent gains in both naturalness and intelligibility after fine tuning. 
In general, these findings suggest that fine-tuning produces perceptible improvements in perceived naturalness across both languages while maintaining intelligibility.


\section{Data Availability Statement}

The dataset described in this study is publicly available to facilitate reproducibility and further research. The data can be accessed via ScienceDB at the following persistent link: [DOI/URL]. The dataset are released under CC BY 4.0, allowing reuse with proper attribution. If you use this dataset, please cite this paper using the following citation information:
Zhitong Zhou, Qingqing Zhang, and Ruohua Zhou. Open-Source Full-Duplex Conversational Datasets for Natural and Interactive Speech Synthesis. Data Intelligence. DOI:

\end{document}